\documentclass{iopconfser}
\usepackage{graphicx}
\usepackage{braket}
\usepackage{color}
\usepackage{amsmath}
\usepackage{float}
\usepackage{amsfonts}
\usepackage{amssymb}
\usepackage{mathtools}
\usepackage{amsthm}
\usepackage{dsfont}
\usepackage{physics}
\usepackage{orcidlink}

\begin{document}

\title{Revival and instabilities of entanglement in monitoring maps with indefinite causal order}

\author{Shahana Aziz$^1$\orcidlink{0000-0002-0567-3541}, André H. A. Malavazi$^1$\orcidlink{0000-0002-0280-0621}, and  Pedro R. Dieguez$^1$\orcidlink{0000-0002-8286-2645}}

\affil{$^1$International Centre for Theory of Quantum Technologies, University of Gdańsk, Jana Bażyńskiego 1A, 80-309 Gdańsk, Poland}

\email{pedro.dieguez@ug.edu.pl}

\begin{abstract}
    In this proceeding, we revisit the discussion presented in Ref. [Commun Phys 7, 373 (2024)], which examines the behavior of a quantum switch involving two arbitrary quantum operations when the control is exposed to environmental effects. Our study extends this analysis by focusing on the evolution of entanglement in the target system within the quantum switch framework, taking into account the influence of environmental conditions and control post-selection. We find that entanglement evolution is highly sensitive to these factors. While entanglement sudden death occurs under definite causal order, indefinite causal order can reverse this loss. We observe entanglement revival in high-temperature regimes and a sudden reappearance of entanglement under weak monitoring conditions at low temperatures. These findings provide insights into the resilience of the quantum switch in the presence of environmental disturbances and highlight its potential for applications where preserving entanglement is essential.
\end{abstract}

\section{Introduction}

In classical physics, causality serves as a cornerstone principle that governs the sequential progression of physical events, where the temporal order between two connected events, A and B, remains well-defined, dictated by a time-like vector. Event A either precedes event B or vice versa, preserving a clear causal relationship. However, this intuitive notion becomes profoundly more intricate within the realm of quantum mechanics. The principle of superposition, foundational to quantum theory, extends beyond states and observables to include causal relations, giving rise to scenarios characterized by indefinite causal order. A compelling manifestation of this phenomenon is observed in quantum processes like the quantum switch (QS), where an auxiliary quantum control system creates a superposition of causal orders (SCO) on a target system. 

At its core, quantum coherence is a distinctive property of the quantum description of nature, fundamentally distinguishing it from classical theories~\cite{baumgratz2014quantifying, streltsov2017colloquium, wu2020quantum, designolle2021set, dieguez2022experimental, giordani2023experimental}.
Building on the concept of superposition of causal orders (SCO), groundbreaking applications have emerged across diverse domains in quantum technologies~\cite{rozema2024experimental}. For instance, SCO has been interpreted as a superposition of time evolution~\cite{felce2022indefinite}, which encourages advances in computation, by enhancing processing capabilities~\cite{chiribella2013quantum, araujo2014computational, procopio2015experimental}, and communication, enabling improved efficiency in information transfer~\cite{ebler2018enhanced, wei2019experimental, guo2020experimental}. In metrology, SCO-based approaches offer unprecedented precision in measurements~\cite{zhao2020quantum}, while in thermodynamics, they facilitate novel approaches to energy management and thermal processes, unlocking new possibilities for quantum heat engines and resource optimization~\cite{felce2020quantum, rubino2021quantum, nie2022experimental, Simonov2022, dieguez2023thermal, Simonov2023}. These applications highlight how harnessing the indefiniteness of causal structures can translate quantum coherence into concrete advantages across various fields~\cite{milz2022resource}.

However, despite the recent theoretical advances, 
it is crucial to account for more realistic scenarios when considering the experimental realizations of the QS. One significant factor that must not be overlooked is the role of environmental degrees of freedom, which are intrinsic to a wide range of matter-based platforms~\cite{felce2021refrigeration, nie2022experimental, RevModPhys.95.045005}.
The complex role played by an environment is central to studies on open quantum systems~\cite{Breuer} and quantum thermodynamics~\cite{Gemmer,binder2018thermodynamics}, and critical to the practical implementation of quantum technologies in general.
Thus, it is desirable to mitigate potentially detrimental effects, such as decoherence~\cite{zurek2003decoherence, schlosshauer2005decoherence}, or to manipulate different energy fluxes to design different thermal devices~\cite{PhysRevE.87.042123,bhattacharjee2021quantum, myers2022quantum, VIEIRA2023100105, PhysRevE.109.064146}.
In this context, given the essential nature of the correlations to achieve SCO~\cite{chiribella2013quantum, goswami2018indefinite}, Molitor et al.~\cite{Molitor2024CommPhys} investigated how an environment affects the operation of the QS, assuming that the control system interacts with an external thermal reservoir. Their work provides a comprehensive framework for understanding and compensating for environmental disturbances, offering practical guidelines for maintaining coherence and optimizing the utility of the QS in realistic and noisy conditions. Such insights are indispensable as the field moves closer to translating the theoretical elegance of quantum causal structures into robust, real-world technologies.

In this work, we focus on the monitoring of two mutually unbiased bases (MUBs) and analyze their behavior under different measurement regimes. Monitoring maps, also known as generalized measurement maps \cite{oreshkov2005weak,dieguez2018information,PhysRevA.111.012220,Molitor2024}, provide a unified framework for describing the transition between weak and strong (projective) nonselective measurements. Weak measurements have proven to be valuable tools in various practical applications~\cite{protect1,protect2,lisboa2022experimental,dieguez2023thermal,VIEIRA2023100105,malavazi2024weak}. Beyond their practical significance, these measurements have also deepened our understanding of fundamental quantum phenomena. In particular, monitoring maps have played a key role in exploring the emergence of realism from the quantum domain \cite{dieguez2018information,PhysRevA.111.012220,mancino2018information,basso2022reality}, and is necessary to define a measure of quantum discord with weak measurements~\cite{dieguez2018weak,lustosa2025emergence}.
Building on these concepts, we investigate a new example not analyzed in Molitor et al. \cite{Molitor2024CommPhys}, focusing on entanglement sudden death in the monitoring of two MUBs with definite order. Specifically, we examine the concurrence of the output state after the sequential monitoring of an initially maximally entangled pair. 
\section{Quantum switch with open control}
Considering a target quantum system, denoted as $S$, characterized by its density matrix $\rho_{S}$ and its temporal evolution governed by the Hamiltonian $H_{S}$, the QS implements a controlled sequence of quantum operations of quantum maps on $S$. The defining attribute of this process lies in the fact that the order in which these quantum maps are applied is conditioned on the quantum state $\rho_{C}$ of an auxiliary control quantum system $C$ , itself associated with Hamiltonian $H_{C}$. A critical requirement is that the dimensionality of the control system $C$ must be sufficient to accommodate the number of quantum maps involved in the operation. 
As described in more detail below, a superposition of causal orders (SCO) can be \textit{locally} realized in the system $S$ provided that a suitable post-selection is performed on the state of the control system. Throughout this work, we adhere to natural units, setting $\hbar = k_B = 1$, for simplicity and coherence in our calculations.

Consider the simplest scenario of a two-level quantum control system. Assuming the initial separability of the $SC$ composite state, and two completely positive trace-preserving (CPTP) maps $\mathcal{M}$ and $\mathcal{N}$ characterized by their respective sets of Kraus operators $\{M_{i}\}$ and $\{N_{j}\}$, which satisfy the completeness relation $\sum_i M_i^\dagger M_i = \sum_i N_j^\dagger N_j = \mathds{1}_S$, the QS generates a quantum state of the form
\begin{equation}
   \rho_{SC}^{\mathcal{M} \leftrightarrow \mathcal{N}} \coloneqq \mathcal{S}_{\mathcal{M},\mathcal{N}}(\rho_S \otimes \rho_C) = \sum_{ij} W_{ij}(\rho_S \otimes \rho_C) W_{ij}^{\dagger},
   \label{eq:switch}
\end{equation}
where $W_{ij} \coloneqq M_i N_j \otimes |0\rangle \langle0|_C + N_j M_i \otimes |1\rangle\langle 1|_C$ is the controlled-Kraus operators of the QS map, with $\{|0\rangle_C,|1\rangle_C\}$ representing the computational basis of $C$.
Thus, the state of the control system $C$ becomes directly linked with the order of application of the maps, with this order ultimately established by a post-selection on the control,
i.e, it is clear that the computational basis elements are explicitly coupled with a \textit{definite} order of application of the maps, such that measurements in this particular basis result in either applying first $\mathcal{N}$ and then $\mathcal{M}$ or vice versa.
However, the situation becomes more intriguing when considering a coherent superposition of causal orders. For instance, when the control system is prepared in a coherent state, such as the eigenstates $\{|+\rangle_C,|-\rangle_C\}$ of the $x$-Pauli operator $\sigma_x$, the resultant quantum state embodies an \textit{indefinite} superposition of causal orders.
To illustrate this, let us assume that the initial state of the control system is given by $\rho_C=|+\rangle \langle+|_C$. Under this condition, we can define two significant operators that emerge from this indefinite causal configuration: 
\begin{equation}
\label{eq:A_def}
    A_{\textrm{def}} \coloneqq \frac{1}{2}\sum_{i,j}\left(M_{i}N_{j}\rho_{S}N_{j}^{\dagger}M_{i}^{\dagger}+N_{j}M_{i}\rho_{S}M_{i}^{\dagger}N_{j}^{\dagger}\right)
\end{equation}
and
\begin{equation}
\label{eq: A_indef}
    A_{\textrm{indef}} \coloneqq \frac{1}{2}\sum_{i,j}\left(M_{i}N_{j}\rho_{S}M_{i}^{\dagger}N_{j}^{\dagger}+N_{j}M_{i}\rho_{S}N_{j}^{\dagger}M_{i}^{\dagger}\right).
\end{equation}
It is important to observe that, while $A_{\textrm{def}}$ represents a convex combination of the maps applied in definite orders, $A_{\textrm{indef}}$ corresponds to a mixture of terms without a definite structure.

Along these lines, Eq.~\eqref{eq:switch} for the post-switch operation simplifies to~\cite{Simonov2023}
\begin{equation}
\label{eq:SCstate_+input}   \rho_{SC}^{\mathcal{M}\leftrightarrow\mathcal{N}}=A_{++}\otimes\ketbra{+}{+}+A_{+-}\otimes\ketbra{+}{-}+A_{-+}\otimes\ketbra{-}{+}+A_{--}\otimes\ketbra{-}{-},
\end{equation}
where $A_{xy} \coloneqq \frac{1}{4}\sum_{i,j}\left[M_{i},N_{j}\right]_{x}\rho_{S}\left[M_{i},N_{j}\right]_{y}^{\dagger}$, with $x,y\in\{+,-\}$ and $[X,Y]_{\pm} \coloneqq XY \pm YX$.
Thus, since
\begin{equation}
\label{eq:ApmpmIntermsofAdefindef}
    A_{\pm\pm}=\frac{1}{2}A_{\rm def}\pm \frac{1}{2}A_{\rm indef},
\end{equation}
one can see that both terms contribute in linearly independent components of Eq.~\eqref{eq:SCstate_+input}. However, this decomposition does not imply the existence of superposition of causal orders and SCO effects. In fact, the local state of $S$ after the QS is given by
\begin{equation}
    \rho_S=\tr_C\left\{\rho_{SC}^{\mathcal{M}\leftrightarrow\mathcal{N}}\right\}=A_{++}+A_{--}=A_{\rm def},
\end{equation}
which contains no features indicative of an SCO.
More importantly, this observation highlights the essential role of measuring the control system.
Indeed, when post-selection is performed in the $\sigma_x$ basis, the probability of each outcome becomes
\begin{equation}
\label{eq:post_prob_UsualSwitch}
    p_{\rm post}(\pm)=\tr_{SC}\left\{\left(\mathds{1}_S \otimes |\pm\rangle \langle\pm|_C\right ) \rho_{SC}^{\mathcal{M} \leftrightarrow \mathcal{N}}\right\} = \tr_S\left\{A_{\pm\pm}\right\},
\end{equation}
resulting in the conditional state
\begin{equation}
    \label{eq:conditionalS_+input}\rho_{S,\pm}^{\mathcal{M}\leftrightarrow\mathcal{N}}= \frac{A_{\pm\pm}}{\tr\left\{A_{\pm\pm}\right\}}.
\end{equation}
Therefore, according to Eq.~\eqref{eq:ApmpmIntermsofAdefindef}, it is clear that the conditional states obtained after the post-selection of $C$ inherit the terms associated with SCO This is precisely why post-selection is indispensable: after measurement and the renormalization procedure, the SCO becomes detectable. In other words, post-selection of the control is a crucial step for revealing SCO effects in the QS mechanism. 

Given the pivotal role that the control system plays in the QS operation, it becomes particularly important to extend the theoretical considerations to more physical and realistic scenarios. In this sense, the analysis in Ref.~\cite{Molitor2024CommPhys} considers the control to interact with an external environment $E$ between the QS map operation and its subsequent post-selection. For simplicity, the influence of the environment is described by the \textit{collisional model} framework\footnote{It is worth mentioning that the collisional model is compatible with the dynamics obtained by the usual GKLS master equation~\cite{gorini1976completely,ciccarello2022quantum}.}, which conceptualizes the environment as a collection of auxiliary qubits, each characterized by a Hamiltonian $H_{E}$ prepared in a quantum state $\rho_E$ sequentially interacting with $C$ via a specified interaction Hamiltonian $V_{CE}$ as discussed in Ref.~\cite{ciccarello2022quantum}.

Each individual interaction, commonly referred to as ``collision'' is assumed to take place over a fixed time duration $\tau$. The combined dynamics during each collision are governed by a unitary operator $U = \text{exp}(-i\tau H_\text{tot})$, where $H_\text{tot} = H_S + H_C + H_{E} + V_{CE}$ is the total Hamiltonian of the composite system. Hence, after $n$ collisions, the joint state $SC$ is written as
\begin{equation}
\label{eq:stateupdatecontrolcoll}
    \rho_{SC}^n = \tr_{E}\left \{ U \left (\rho_{SC}^{n-1} \otimes \rho_{E} \right ) U^\dagger \right \}.
\end{equation}
Naturally, the precise nature of these dynamics depends on the specifics of each model considered. Here, we examine the case the environment behaves as a thermal reservoir with temperature $T_E$. Accordingly, each auxiliary qubit in the reservoir is expressed by a Gibbs state such that
\begin{equation}
    \rho_{E}=\Theta_{E} = \frac{\text{exp}(-\beta_E H_{E})}{Z_{E}},
\end{equation}
where $Z_{E} = \tr\{\text{exp}(-\beta_E H_{E})\}$ is the partition function and $\beta_E=1/T_E$ is the inverse of temperature.
The interaction between the control system and the environmental qubit is modelled by the Hamiltonian
\begin{equation}\label{interaction}
    V_{CE} = \frac{g}{2} \left( \sigma_z^C \sigma_z^{E} + \sigma_y^C\sigma_y^{E}  \right)=g\left(|+\rangle \langle-|_{C}\otimes |-\rangle \langle +|_{E}+\text{h.c.}\right),
\end{equation}
where $g$ denotes the coupling strength between the control and the environment. The collisions are treated as an individual perturbation in the control state, being fast relative to the relevant temporal scale, s.t. $g\tau\ll 1$.
This interaction Eq.~\eqref{interaction} has an isotropic structure and is equivalent to the standard Jaynes-Cummings coupling for a reservoir of qubits~\cite{Ciccarello_2013}, describing the exchange of excitation \cite{cusumano2022quantum, ciccarello2022quantum}.
Additionally, without loss of generality, we fix $H_{C, E} = -\omega \sigma_x^{C, E}/2$, which guarantees the post-selection basis of the control system, $\{\ket{+}_{C},\ket{-}_{C}\}$, is invariant over its free time-evolution (up to a phase).
From a thermodynamic perspective, the chosen model guarantees that no work is performed due to the interaction~\cite{Molitor2020} and that each collision is consistent with a thermal operation~\cite{PhysRevLett.111.250404, Lostaglio2018}, satisfying the strict energy conservation condition for each energy exchange~\cite{dann2021open}, i.e., $[H_C + H_{E}, V_{CE}]_- = 0$.

Therefore, building upon Eq.~\eqref{eq:SCstate_+input}, one can analytically compute the joint $SC$-state after $n$ successive collisions with the environment. The resulting state is given by
\begin{equation}
\label{eq:GenExpAfterColl}   
        \rho_{SC}^{n}= \mathcal{B}_{++}(n)\otimes|+\rangle\langle+|_C+\mathcal{B}_{+-}(n)\otimes|+\rangle \langle-|_C
        +\mathcal{B}_{-+}(n)\otimes|-\rangle \langle+|_C+\mathcal{B}_{--}(n)\otimes|-\rangle \langle-|_C,
\end{equation}
where the coefficients are defined as:
$\mathcal{B}_{+-}(n)=\mathcal{B}_{-+}^{\dagger}(n)\coloneqq e^{in\tau\omega}\cos^{n}(\text{g\ensuremath{\tau}})U_{S}^{n}A_{+-}U_{S}^{\dagger n}$ and
\begin{equation}\label{eq:Bpmpm}
    \mathcal{B}_{\pm\pm}(n)\coloneqq\frac{b^{\pm}_{\rm def}(n,f_E,g\tau)}{2}A_{\rm def}^n + \frac{b^{\pm}_{\rm indef}(n,g\tau)}{2}A^n_{\rm indef},
\end{equation}
where $U_{S}\coloneqq\text{exp}(-i \tau H_S)$ represents the time-evolution operator of the target system $S$, and we defined $f_E \coloneqq \tanh (\beta_E \omega/2)$, $A_{\text{def}}^{n}\coloneqq U_{S}^{n}A_{\textrm{def}}U_{S}^{\dagger n}$,  $A_{\text{indef}}^{n}\coloneqq U_{S}^{n}A_{\textrm{indef}}U_{S}^{\dagger n}$, and
\begin{equation}
    \label{eq:bdef&indef}    
        b^{\pm}_{\rm def}(n,f_E,g\tau) := 1\pm f_{E}\left[1-\cos^{2n}(\text{g\ensuremath{\tau}})\right], \qquad
        b^{\pm}_{\rm indef}(n,g\tau) :=\pm\cos^{2n}(\text{g\ensuremath{\tau}}).
\end{equation}
It is important to emphasize that Eq.~\eqref{eq:GenExpAfterColl} is completely general for any 2-quantum switch, maintaining an analogous form to Eq.~\eqref{eq:SCstate_+input}. Hence, after the post-selection, the probability of obtaining a particular control state is
\begin{align}
\label{eq:postselectionprobBpmpm}
    p^n_{\rm post}(\pm)=\tr\left\{\left(\mathds{1}_S \otimes |\pm\rangle \langle\pm|_C\right ) \rho_{SC}^{n}\right\} = \tr\left\{\mathcal{B}_{\pm\pm}(n)\right\},
\end{align}
and the corresponding conditional state of the target system becomes
\begin{align}
        \label{eq:conditionalStateBpmpm}
        \rho_{S,\pm}^{n}=\frac{\mathcal{B}_{\pm\pm}(n)}{p^n_{\rm post}(\pm)}.
\end{align}

We highlight the term $b_{\rm indef}^\pm$ characterizes the contribution of SCO within the quantum switch. A clear observation is that environmental collisions lead to a monotonic suppression of SCO effects.
To illustrate the environmental role, it is instructive to analyze limiting scenarios. In the asymptotic limit of infinitely many collisions, $n\rightarrow \infty$, the joint state takes the form
\begin{equation}\label{eq: assymptoticSCstate}
    \lim_{n\to\infty}\rho_{SC}^n = \left[\lim_{n\to\infty}\mathcal{B}_{++}(n)\right]\otimes|+\rangle \langle+|_C + \left[\lim_{n\to\infty}\mathcal{B}_{--}(n)\right]\otimes |-\rangle \langle-|_C=\rho^\infty_S\otimes\Theta_{\beta_{E}},
\end{equation}
where $\rho_{S}^{n}=A_{\text{def}}^{n}$, i.e. the correlations are suppressed, and the joint system converges to a product state between the time-evolved mixture of the causally ordered contribution of $S$ and the thermal state of the control relative to the environment's temperature. On the other hand, for $n=0$ the Eq.~\eqref{eq:SCstate_+input} for a closed control system is naturally recovered, reflecting the absence of any environmental influence.  
The impact of temperature $T_E$ on the QS as $n\to\infty$ reveals distinct behaviors.
At high temperatures, the SCO contributions for both post-selected states are significantly suppressed, due to stronger thermal fluctuations, with the probabilities of obtaining either post-selected control state approaching the uniform value of $1/2$. In contrast, at low temperatures, a more nuanced behavior emerges: while the SCO contribution for the post-selected state, $\rho^n_{S,+}$ remains suppressed, the contribution associated with the post-selected state, $\rho^n_{S,-}$, appears to be remarkably shielded from environmental interactions, effectively rendering it independent of the number of collisions. However, despite the apparent shielding, the probabilities of post-selecting $\ket{-}_C$ and $\ket{+}_C$  asymptotically approach to $0$ and $1$, respectively, reflecting a thermally driven asymmetry in the post selection process as the system evolves toward its steady-state configuration~\cite{Molitor2024CommPhys}.
\section{Entanglement sudden-death with definite order, and entanglement sudden-revival with unstable indefinite causal order}

Monitoring maps are CPTP maps that interpolate between weak and strong non-selective measurements. Assuming that $\rho_S$  is an initial bipartite state $\rho_{AB}$, the monitoring map acting on subsystem $A$ results in~\cite{oreshkov2005weak, dieguez2018information}
\begin{equation}
    \mathcal{M}_{\mathcal{O}}^{\epsilon}(\rho_{AB}) \coloneqq (1-\epsilon) \rho_{AB} + \epsilon\, \Phi_{\mathcal{O}}(\rho_{AB}),
    \label{eq:mon-map}
\end{equation}
where $0\leqslant\epsilon\leqslant1$ is the measurement strength and the map $\Phi_{\mathcal{O}}$ is a dephasing of subsystem $A$ in the eigenbasis of the operator $\mathcal{O}=\sum_\alpha \alpha \, \mathcal{O}_\alpha$, i.e.,
\begin{equation}
    \label{eq:reality}
    \Phi_\mathcal{O}(\rho_{AB}) \coloneqq \sum_{\alpha} (\mathcal{O}_\alpha\otimes \mathds{1}_{B})  \rho_{AB}  (\mathcal{O}_\alpha\otimes \mathds{1}_{B}) =\sum_\alpha p_\alpha \mathcal{O}_\alpha\otimes \rho_{B|\alpha}
\end{equation}
where $\rho_{B|\alpha} = \bra{\alpha} \hat{\rho} \ket{\alpha}/p_{\alpha}$ with $p_\alpha = \tr\{(\mathcal{O}_\alpha \otimes \mathds{1})\rho_{AB}\}$, and $\mathcal{O}_\alpha$ are projectors such that $\mathcal{O}_\alpha \mathcal{O}_{\alpha'}=\delta_{\alpha \alpha'}\mathcal{O}_\alpha$. Map $\Phi_{\mathcal{O}}$ can be interpreted as the projective measurement of observable $\mathcal{O}$ without having its outcome revealed. A possible choice of Kraus decomposition for this operation is given by $K_0 = \sqrt{1-\epsilon} \,\mathds{1}_{A}\otimes \mathds{1}_B$ and $K_j = \sqrt{\epsilon}\, \mathcal{O}_j\otimes \mathds{1}_B$. 

Now, let us analyze a simple application of the results introduced in Ref.~\cite{Molitor2024CommPhys} and previously reviewed here for two qubits in an initial maximally entangled state.

\subsection{Bell pair}
Concurrence $\mathcal{C}$ is a measure of quantum entanglement, ranging from $0$ (no entanglement) to $1$ (maximally entangled).
For a generic two-qubit pure state $\rho_{AB}=\ket{\psi}\bra{\psi}$ with $ |\psi\rangle = \alpha|00\rangle + \beta|01\rangle + \gamma|10\rangle + \delta|11\rangle $ and  $\left\{|0(1)\rangle\right\}$ being the local computational basis, this quantity is defined as
$\mathcal{C}_{A(B)} = \sqrt{2 \left( 1 - \tr_{A(B)}\left\{\rho^2\right\} \right)}$,
where $\tr\left\{\rho_{A(B)}^{2}\right\}$ is the purity of the local state $\rho_{A(B)}$.
In contrast, for a mixed state the concurrence is given by
$\mathcal{C} = \max\left(0, \lambda_1 - \lambda_2 - \lambda_3 - \lambda_4\right)$,
where $\left\{\lambda_j \right\}_{j=1,...,4}$ are the eigenvalues, in decreasing order, of
\begin{equation}
R = \sqrt{\sqrt{\rho} (\sigma_y \otimes \sigma_y) \rho^* (\sigma_y \otimes \sigma_y)\sqrt{\rho}},
\end{equation}
with $\sigma_y $ being the Pauli-Y matrix and $\rho^{*}$ the complex conjugate of the density matrix $\rho$.

Now, let us consider a Bell state $\rho_{AB}=\ket{\varphi}\bra{\varphi}$ where $\ket{\varphi}=\frac{1}{\sqrt{2}}(\ket{00}+\ket{11})$ and apply the QS consisting of two MUBs monitoring.
We start by performing the ideal quantum switch, and then showing how to apply the open control condition by mapping the operators $A_{\pm\pm}$ to $\mathcal{B}_{\pm\pm}$.
First, we explicitly compute the following operators $A_{\pm\pm}$ considering the initial Bell state:
\begin{equation}
    A_{++} = \frac{1}{4}\begin{bmatrix}

\begin{array}{cccc}
 2-\epsilon  & 0 & 0 &  (\epsilon -2) (\epsilon -1) \\
 0 & -\frac{1}{2} (\epsilon -2) \epsilon  & -\frac{1}{2} (\epsilon -2) \epsilon  & 0 \\
 0 & -\frac{1}{2} (\epsilon -2) \epsilon  & -\frac{1}{2} (\epsilon -2) \epsilon  & 0 \\
  (\epsilon -2) (\epsilon -1) & 0 & 0 & 2-\epsilon \\
\end{array}
    \end{bmatrix}
\end{equation}
and
\begin{equation}
    A_{--} = \frac{1}{8}\begin{bmatrix}
\begin{array}{cccc}
 0 & 0 & 0 & 0 \\
 0 & \epsilon ^2 & -\epsilon ^2 & 0 \\
 0 & -\epsilon ^2 & \epsilon ^2 & 0 \\
 0 & 0 & 0 & 0 \\
\end{array}
    \end{bmatrix}
\end{equation}
The respective probabilities and resulting conditional states are given by $ p_{\rm post}(+) = 1-\frac{\epsilon ^2}{4}$ and $p_{\rm post}(-) = \frac{\epsilon ^2}{4}$, and
\begin{equation}
    \rho_{AB,+} = \frac{A_{++}}{\tr\left\{ A_{++}\right\}} = \begin{bmatrix}
\begin{array}{cccc}
 \frac{1}{\epsilon +2} & 0 & 0 & \frac{3}{\epsilon +2}-1 \\
 0 & \frac{\epsilon }{2 \epsilon +4} & \frac{\epsilon }{2 \epsilon +4} & 0 \\
 0 & \frac{\epsilon }{2 \epsilon +4} & \frac{\epsilon }{2 \epsilon +4} & 0 \\
 \frac{3}{\epsilon +2}-1 & 0 & 0 & \frac{1}{\epsilon +2} \\
\end{array}
    \end{bmatrix}
\end{equation}
and 
\begin{equation}
    \rho_{AB,-} = \frac{A_{--}}{\tr\left\{ A_{--}\right\}} = \frac{1}2{}\begin{bmatrix}
\begin{array}{cccc}
 0 & 0 & 0 & 0 \\
 0 & 1 & -1 & 0 \\
 0 & -1 & 1 & 0 \\
 0 & 0 & 0 & 0 \\
\end{array}
    \end{bmatrix},
\end{equation}
which is the Bell state, $\rho_{AB}^{\prime} = |\Psi^{-}\rangle\langle\Psi^{-}|$, where $|\Psi^{-}\rangle = \frac{1}{\sqrt{2}}(|10\rangle-|01\rangle)$. 
\begin{figure}[t]
\centering
\includegraphics[width=1\columnwidth]{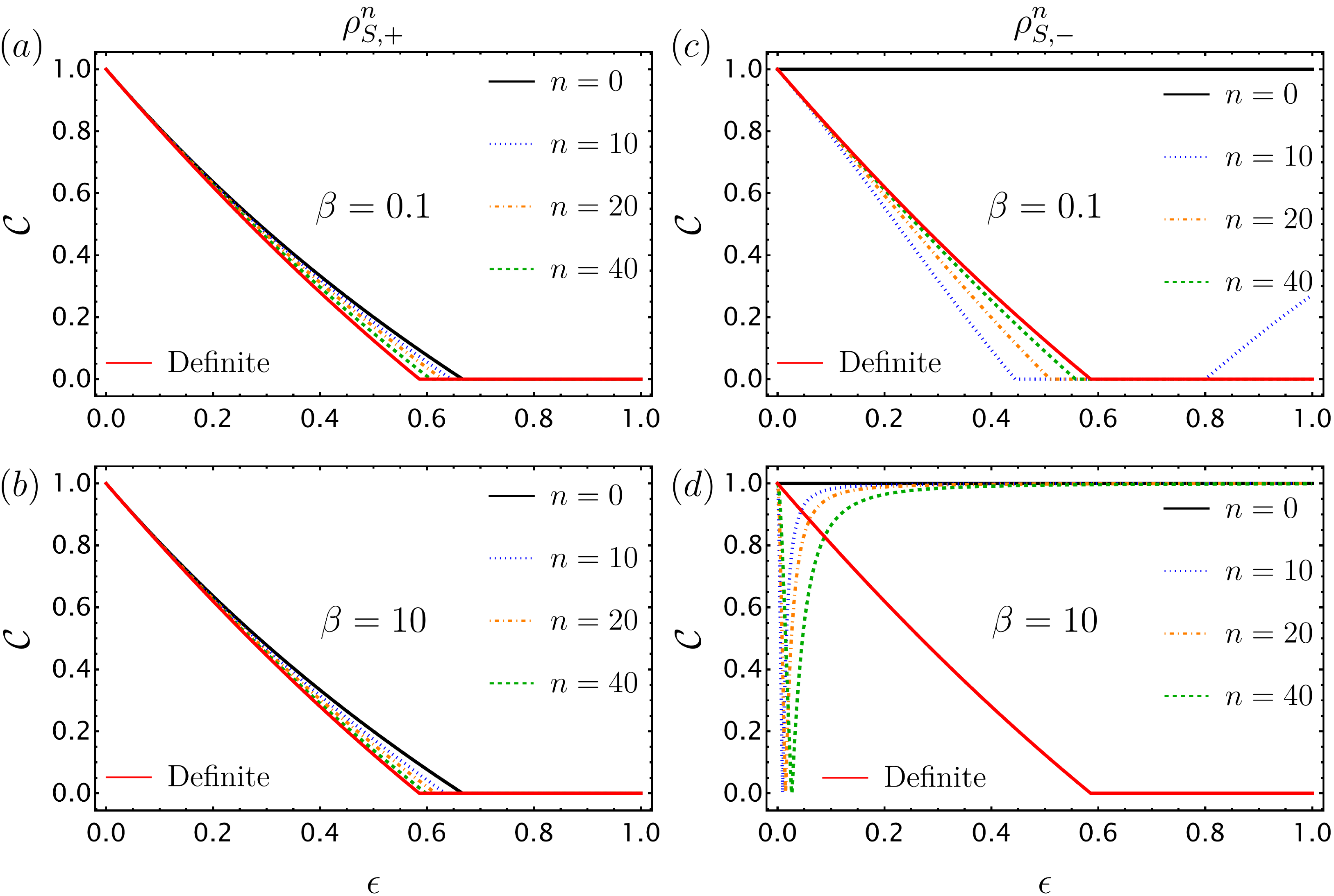}
\caption{\textbf{Concurrence after the post-selection of the control as a function of the measuring strength.} Here, we assume $\omega_{S}=\omega=1$, and $g\tau=0.2$, and $\beta$ has units with the inverse of $\omega_S$. We consider the behavior of $\mathcal{C}(\rho_{AB,\pm}^n)$ as a function of the strength $\epsilon$ of the monitoring maps for different numbers of collisions $n$ and for both post-selected states via the $\ket{\pm}$-control measurement. It is considered the cases in which the control has been exposed to an environment with (a)-(c) high and (b)-(d) low temperatures. The red curve (``definite'') shows the concurrence for the monitoring maps applied in definite causal order.}
\label{fig1}
\end{figure}

Now, we can proceed and write $\mathcal{B}_{\pm \pm}$ in terms of the previous operators.
Using
\begin{center}
\begin{equation}
  A_{++}+A_{--} = \frac{1}{4}\begin{bmatrix}
\begin{array}{cccc}
 2-\epsilon  & 0 & 0 &  (\epsilon -2) (\epsilon -1) \\
 0 & \epsilon  & \epsilon(1-\epsilon)  & 0 \\
 0 & \epsilon(1-\epsilon)  & \epsilon  & 0 \\
  (\epsilon -2) (\epsilon -1) & 0 & 0 & 2-\epsilon 
\end{array}
\end{bmatrix}
\end{equation}
\end{center}
and
\begin{center}
\begin{equation}
  A_{++}-A_{--} = \frac{1}{4}\begin{bmatrix}
     
\begin{array}{cccc}
 2-\epsilon & 0 & 0 &  (\epsilon -2) (\epsilon -1) \\
 0 & \epsilon(1-\epsilon)  & \epsilon  & 0 \\
 0 & \epsilon  & \epsilon(1-\epsilon) & 0 \\
 (\epsilon -2) (\epsilon -1) & 0 & 0 & 2-\epsilon  \\
\end{array}
 \end{bmatrix},
 \end{equation}
\end{center}
and assuming the appropriated post-selected state after the QS, we use the expressions given by Eqs.~\ref{eq:Bpmpm} and~\ref{eq:bdef&indef} to get the open control result after $n$ collisions as
\begin{equation}
    \mathcal{B}_{\pm\pm}(n)=\frac{1\pm f_{E}[1-\cos^{2n}]}{2}U_{AB}^{n}(A_{++}+A_{--})U_{AB}^{\dagger n} + \frac{\pm\cos^{2n}(\text{g\ensuremath{\tau}})}{2}U_{AB}^{n}(A_{++}-A_{--})U_{AB}^{\dagger n},
\end{equation}
where $U_{AB}$ is the free evolution of the bipartition, which-without loss of generality-we will assume to be $H_{AB}=\sigma_z\otimes \mathds{1}_{B}+\mathds{1}_{A}\otimes\sigma_z$. 

With that, we can thoroughly compute the post-selected states $\rho_{AB,\pm}^{n}=\mathcal{B}_{\pm\pm}(n)/p^n_{\rm post}(\pm)$ in terms of $n$.
Hence, by plotting the concurrence for both post-selections with an open control, we can analyze how environmental temperature influences the instabilities in each case. This result is illustrated in Fig.~\ref{fig1}, with the red curves representing the concurrence behavior for the monitoring maps applied in definite causal order.
Our results reveal that for the control post-selection realized by $\ket{+}$, entanglement sudden death persists. However, in this case, the interference of causal order slightly increases the measurement strength $\epsilon$ required for sudden death in both high- and low-temperature regimes, with the curves approaching the definite one for higher values of $n$. 
The behavior for the $\ket{-}$ post-selection, however, is considerably different. In the high-temperature regime, entanglement may revive after its initial death depending on $n$, while in the low-temperature regime, decay is followed by a sudden revival at weak monitoring strengths. Moreover, as the measurement strength increases, the result gradually recovers the ideal quantum switch behavior. 
Notice that, in the ideal case for this projection, entanglement is always recovered when employing the quantum switch, independently of the collisions. 
These results highlight the intricate role of indefinite causal order in mitigating entanglement loss under different environmental conditions~\cite{oreshkov2005weak, dieguez2018information, dieguez2018weak}.   
\section{Discussion} 

In this work, we have reviewed the open control switch model introduced in Ref.~\cite{Molitor2024CommPhys}, to study how environmental interactions affect superposed causal order (SCO) under two arbitrary quantum operations. This framework highlights the role of the control system in mediating instabilities and shaping the effectiveness of indefinite causal orders.  
 By incorporating two mutually unbiased basis (MUB) measurements within the quantum switch, we analyzed how the switch and its instabilities influence entanglement dynamics, quantified by concurrence, in the target system. 
 
 Our example shows that entanglement on the target system critically depends both on the control post-selection and environmental conditions. While entanglement sudden death still occurs under certain post-selection, indefinite causal order effects can protect or even reverse this loss. More importantly, we observe an entanglement revival in high-temperature regimes and a sudden reappearance at weak monitoring strengths in low-temperature conditions. In the latter, as the measurement strength increases, the system gradually restores the ideal quantum-switch behavior, ensuring entanglement preservation.  
These results provide insight into the quantum switch and its resilience to uncontrolled environmental noise. The observed revival of entanglement suggests new possibilities for using indefinite causal order in quantum information processing, especially in applications where maintaining coherence and entanglement is essential.

Future research should further investigate how the environment interacts with different control-system configurations to maximize the benefits of indefinite order.
One possible approach to mitigating unwanted noise effects is through weak measurements, which have been shown to help preserve coherence in various scenarios despite environmental interference~\cite{kim2012protecting,Xiao2013,protect1,protect2,protect3,protect4,Lee:11,Lalita_2024,malavazi2024weak}.
Additionally, these instabilities could affect any quantum-controlled protocol that requires a final measurement in the control system, making it vulnerable to environmental disturbances. This issue is particularly relevant for protocols involving the superposition of operations~\cite{ebler2018enhanced, abbott2020communication}, the superposition of opposite time directions~\cite{chiribella2022quantum}, and quantum-controlled delayed-choice experiments~\cite{dieguez2022experimental}.
\section*{Acknowledgements}
S.A and P.R.D. acknowledge the support from the Narodowe Centrum Nauki (NCN) Poland, ChistEra-2023/05/Y/ST2/00005 under the project Modern Device Independent Cryptography (MoDIC).
A.H.A.M. acknowledges support from the NCN, Poland Grant OPUS-21 (No. 2021/41/B/ST2/03207). 
\bibliography{ref}

\end{document}